\documentclass[fleqn,twoside]{article}
\usepackage{espcrc2}
\usepackage{graphicx}

\def\rme{{\rm e}}

\def\bs{{\tt\char"5C}}

\def\proof{\noindent{\sl Proof:}\kern0.6em}

\def\frac#1#2{\hbox{$#1\over#2$}}
\def\dual{\mathstrut^*\kern-0.1em}

\def\slash#1{\setbox0=\hbox{$#1$}
    \setbox1=\hbox{$/$}
    #1\kern-\wd0\kern1.5pt/\kern-\wd1\kern\wd0}
\def\lvec#1{\setbox0=\hbox{$#1$}
    \setbox1=\hbox{$\scriptstyle\leftarrow$}
    #1\kern-\wd0\smash{
    \raise\ht0\hbox{$\raise1pt\hbox{$\scriptstyle\leftarrow$}$}}
    \kern-\wd1\kern\wd0}
\def\rvec#1{\setbox0=\hbox{$#1$}
    \setbox1=\hbox{$\scriptstyle\rightarrow$}
    #1\kern-\wd0\smash{
    \raise\ht0\hbox{$\raise1pt\hbox{$\scriptstyle\rightarrow$}$}}
    \kern-\wd1\kern\wd0}
\def\boxit#1{\vbox{\hrule height2pt\hbox{\vrule width2pt
    \kern10pt\vbox{\kern10pt#1\kern10pt}\kern10pt\vrule width2pt}
    \hrule height2pt}}

\def\nabstar#1{\nabla\kern2.0pt\smash{\raise8.0pt\hbox{$\ast$}}
               \kern-12.0pt_{#1}\kern3pt}

\def\drvstar#1{\partial\kern2.0pt\smash{\raise8.0pt\hbox{$\ast$}}
               \kern-12.0pt_{#1}\kern3pt}

\def\GeV{{\rm GeV}}

\def\fm{{\rm fm}}

\def\diracstar#1#2{
    \setbox0=\hbox{$\gamma$}\setbox1=\hbox{$\gamma_{#1}$}
    \gamma_{#1}\kern-\wd1\kern\wd0
    \smash{\raise4.5pt\hbox{$\scriptstyle#2$}}}

\def\deltaA#1{\delta_{#1}\kern-1.0pt A}

\title{
       \vspace{-4.0cm}
       \rightline{\normalsize CERN-TH/2001-262}
       \vspace{-0.1cm}
       \rightline{\normalsize October 2001}
       \vspace{2.9cm}
Lattice QCD on PCs\kern1pt?\,%
       \thanks{Plenary talk,
       XIX International Symposium on Lattice Field Theory, 
       August 19--24, 2001, Berlin, Germany}
}

\author{M.~L\"uscher\,%
        \thanks{On leave from Deutsches Elektronen-Synchrotron DESY,
        D-22603 Hamburg, Germany}\\[1.9ex]
        {CERN, Theory Division, CH-1211 Geneva 23, Switzerland}}
       
\begin{document}

\begin{abstract}
Current PC processors are equipped with vector processing units
and have other advanced features that can be used to accelerate
lattice QCD programs. Clusters of PCs with a high-bandwidth network thus 
become powerful and cost-effective machines for numerical simulations.

\vspace{1pc}
\end{abstract}

\maketitle

\section{INTRODUCTION}

Parallel computers built from PC components
are being increasingly used in many branches of science.
The obvious advantages of such machines are that 
the hardware is relatively cheap and that the 
software environment ({\tt linux} operating system, 
the {\tt gcc} compiler suite and an implementation of 
the {\tt MPI} communication library) complies with the established standards.

Some doubts have however been raised that 
PC clusters are good machines for numerical simulations
of lattice QCD.
According to the benchmark results presented at last year's 
lattice conference \cite{GottliebI},
PC processors in fact appear to perform rather 
poorly in these calculations, parti\-cu\-larly so when the lattices
get reasonably big.
Moreover, it remains unclear
how well such clusters scale to large numbers of processors,
where heat dissipation, component reliability and 
the network performance become limiting factors.

The main message of this talk is that
lattice QCD programs can be accelerated by a large factor
if use is made of the vector arithmetic unit 
and other enhancements of current PC processors.
As a consequence (and also for various other reasons) 
the prospects for doing numerical simulations on PC clusters
are now much brighter than they were only a year ago.

\vfill

\section{PC PROCESSOR PERFORMANCE}

\subsection{Multimedia extensions}

The vector processing capabilites mentioned above 
have been added to the recent generations of PC processors
to speed up multimedia applications.
Depending on the brand and type of processor,
the associated instruction sets go under the name of 
{\tt MMX}, {\tt 3DNow!\kern-1pt}, {\tt SSE} and {\tt SSE2}.
In most cases these instructions operate
on short vectors of data in parallel and require
one or two machine cycles to complete. 

Evidently the number of arithmetic operations per cycle that can be
performed also depends on the rate at which the data can be
moved between the memory and the arithmetic units.
To reduce the associated latencies,
current processors support memory-to-cache prefetch
instructions and streaming memory access modes, and they include 
a second-level cache memory that is clocked at the processor
frequency.

\subsection{Benchmark results}

Most of the computer time in numerical simulations of lattice
QCD is spent on the solution of the Dirac equation in the 
presence of a given background gauge field.
There are different ways to solve the equation, but 
in all cases the program that applies the lattice
Dirac operator to a prescribed fermion field
is the one that dominates the execution time.
The rest of the time is used for 
linear combinations and scalar products of Dirac fields.

\begin{table*}[htb]
\caption{Benchmark results obtained with a $1.4$~GHz
Pentium 4 processor}
\label{table:1}
\begin{tabular}{*{5}{l}}
\hline
\noalign{\vspace{1.0ex}}
&\multicolumn{1}{c}{$(D_{\rm w}+m_0)\psi$}
&\multicolumn{1}{c}{$\psi_1+r\psi_2$}
&\multicolumn{1}{c}{$\|\psi\|^2$}
&\multicolumn{1}{c}{CG}\\[1.0ex]
\hline
\noalign{\vspace{1.0ex}}
32-bit\hspace{0.2cm}
& \hspace{0.2cm}0.93 [1503]  
& \hspace{0.3cm}0.13 [363]\hspace{0.3cm}
& \hspace{0.1cm}0.046 [1042]\hspace{0.2cm}
& \hspace{0.1cm}2.34 [1292]
\\[1.5ex]
64-bit
& \hspace{0.2cm}1.71 [814]
& \hspace{0.3cm}0.26 [181]
& \hspace{0.1cm}0.097 [497]
& \hspace{0.1cm}4.40 [687]
\\[1.0ex] 
\hline
\end{tabular}
\\[1.0ex]
(Execution times in $\mu$s per lattice point [speed in Mflop/s])
\vspace{-0.1cm}
\end{table*}

In table~1 some benchmark results 
for these basic programs are reported,
for the case of a PC with $1.4$ GHz Intel Pentium~4 processor
and $256$ MB of PC800 RDRAM.
The two lines in this table contain the processor times required
for the specified task using 32-bit or
64-bit arithmetic. 
In the second column, for example, 
the times for the application
of the Wilson--Dirac operator $D_{\rm w}+m_0$ to a given field
$\psi$ are listed.

All figures quoted in the table are for a $16^4$~lattice,
but on larger lattices they would be practically the same,
because the memory latencies in these programs are effectively
masked by the use 
of memory-to-cache prefetch instructions and 
a cache-optimized data layout.
To accelerate the floating-point arithmetic,
extensive use has been made of the {\tt SSE2} vector instructions
that are supported on the Pentium 4 processor.

Compared with the program for the Dirac operator, the linear
algebra routines (columns~3 and 4 in table~1) appear to be 
rather slow. This can be explained by noting that
these programs spend most of the time to move data from
the memory to the processing units. In a typical QCD code 
the linear algebra programs fortunately consume 
only a fraction of the total time so that their influence
on the overall performance of the code is limited.
This is illustrated by the figures
for a standard conjugate gradient iteration
quoted in the last column of table~1.

The bottom line then is that, with the current generation of 
PC processors and if use is made of their advanced features, 
it is possible to achieve sustained computational
speeds that are about 10 times higher than those quoted
last year \cite{GottliebI}. 

In the remainder of this section, the 
vector unit of the Pentium~4 processor and its usage are discussed
in some detail. 
Another issue that will be addressed is the cache management.
To a large extent the impressive performance of current PC processors
is in fact due to improvements of the memory system.

\subsection{Streaming SIMD extension ({\tt SSE})}

The vector unit on the recent
Intel processors (Pentium III, Pentium 4, Itanium)
has 8 registers that are denoted by {\tt xmm0,...,xmm7}.
They are 128 bits wide and can accommodate
2 double-precision or 4 single-precision
floating-point numbers. 
The associated machine instructions 
operate on these numbers in parallel, 
i.e.~the vector unit 
is a Single Instruction Multiple Data (SIMD) machine 
(see fig.~1).

\begin{figure}[h]
\vspace{-0.35cm}
\hspace{1.3cm}\includegraphics[scale=0.5]{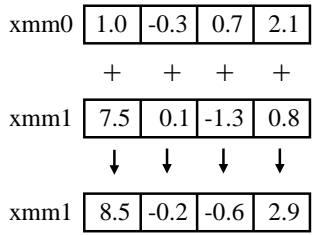}
\vspace{-0.60cm}
\caption{The {\tt SSE} instruction {\tt addps xmm0,xmm1} adds the 
single-precision numbers in register {\tt xmm0} to those 
in {\tt xmm1}.} 
\label{fig:Registers}
\vspace{-0.5cm}
\end{figure}

All basic arithmetic operations (parallel addition, subtraction, multiplication
and division) are supported
and there are further instructions for data moving and shuffling.
On the Pentium III only
single-precision numbers can be manipulated in this way,
but the Pentium~4 supports an extended instruction set,
denoted by {\tt SSE2}, that allows parallel 
double-precision and integer arithmetic
to be performed in the {\tt SSE} registers. 
In all cases
the IEEE-754 standard is respected, which guarantees proper rounding and
exception hand\-ling.

\subsection{Programming example}

When an {\tt ISO}~{\tt C} compliant program is compiled with the {\tt gcc}
compiler, the generated object code will contain general purpose and 
{\tt x87} {\tt FPU} instructions only.
The {\tt SSE} unit is thus not used.
There is, however, a simple way to access the {\tt SSE} registers
from a {\tt C} program and to operate on them. Let us consider
the loop

\vspace{2ex}
{\tt
\noindent
for (i=0;i<100;i++) \\[0.2ex]
\hbox{\hskip0.5cm}a[i]=b[i]+c[i];
}

\vspace{2ex}
\noindent
where {\tt a}, {\tt b}, {\tt c} are assumed to be arrays of 
single-precision floating-point numbers.
Since there are no data dependences in this loop,
we may use the {\tt addps} instruction described above
to do four additions at once.
The code then becomes

\vspace{2ex}
{\tt
\noindent
for (i=0;i<100;i+=4) \\[0.2ex]
\hbox{\hskip0.5cm}\_\_asm\_\_ \_\_volatile\_\_ (\\[0.2ex]
\hbox{\hskip1.0cm}"movaps \%1, \%\%xmm0 \bs n\bs t" \\[0.2ex]
\hbox{\hskip1.0cm}"movaps \%2, \%\%xmm1 \bs n\bs t" \\[0.2ex]
\hbox{\hskip1.0cm}"addps \%\%xmm0, \%\%xmm1 \bs n\bs t" \\[0.2ex]
\hbox{\hskip1.0cm}"movaps \%\%xmm1, \%0" \\[0.2ex]
\hbox{\hskip1.0cm}: \\[0.2ex]
\hbox{\hskip1.0cm}"=m" (a[i]) \\[0.2ex]
\hbox{\hskip1.0cm}: \\[0.2ex]
\hbox{\hskip1.0cm}"m" (b[i]), \\[0.2ex]
\hbox{\hskip1.0cm}"m" (c[i]));
}

\vspace{2ex}
\noindent
The {\tt gcc} compiler understands these lines, and if they
are substituted in any {\tt C} program, the executable 
will perform all 100 additions correctly.

The syntax of the {\em inline assembly} statement {\tt \_\_asm\_\_} 
{\tt \_\_volatile\_\_ (..)}
is documented in the manual pages of the {\tt gcc} compiler.
It has three arguments in the present case,
{\tt a[i]}, {\tt b[i]} and {\tt c[i]}, 
which are labelled from {\tt 0} to {\tt 2}.
The first line {\tt "movaps ..."} moves
four consecutive floating-point numbers from the address of argument 1
to the {\tt SSE} register {\tt xmm0}. The register then contains
the array elements {\tt b[i],...,b[i+3]}. Similarly
the second line moves {\tt c[i],...,c[i+3]} to {\tt xmm1}.
The parallel addition is then performed and the result
is finally moved from {\tt xmm1} 
to the array ele\-ments {\tt a[i],...,a[i+3]}.

This example is rather trivial, but 
more complicated code sequences can be rewritten in essentially 
the same way, with perhaps several inline assembly statements in a row.
Apart from those mentioned above,
many more vector instructions are supported by the Pentium 4 processor
that make it possible 
to perform almost any calculation in the {\tt SSE} registers.
The complete list of {\tt SSE2} instructions, together with 
detailed explanations of what precisely they do, can be downloaded 
from the Intel web pages \cite{Intel}.

In lattice QCD the SU(3) matrix times vector multiplication
is one of the basic operations that is worth being 
coded for the {\tt SSE} unit. On the Pentium 4 processor an average
throughput of $1.8$ single-precision 
floating-point operations per cycle can been achieved in this case.
The perfor\-mance of the double-precision code is equally impressive and
reaches $0.9$ operations per cycle.

\subsection{Cache management}

A common limitation of current
computer architectures is that 
the memory system is often unable to 
deliver the data to the processor at a sufficiently high rate
to keep the arithmetic units busy.
The cache memory (which is significantly faster than the main memory) 
serves to enhance the data availability 
in those cases where the same data are used more than once. 

In the program that applies the Wilson--Dirac operator to a given
fermion field, for example, each spinor of the input field
is processed 9 times when the program runs through the lattice.
So if the spinor components that are needed at the current lattice point
are already in the cache memory (because they have previously been used),
it takes much less time to forward them to the arithmetic units 
than would otherwise be required.

\begin{table*}[htb]
\caption{Cache line sizes [byte] 
and memory-to-processor bandwidths [Gbyte/s]}
\label{table:2}
\begin{tabular}{lcrcc}
\hline
\noalign{\vspace{1.0ex}}
&&\multicolumn{1}{c}{Cache}
&\multicolumn{1}{c}{Theoretical}
&\multicolumn{1}{c}{Measured}
\\[-0.1ex]
\multicolumn{1}{c}{Processor}
&\multicolumn{1}{c}{Memory type}
&\multicolumn{1}{c}{line size}
&\multicolumn{1}{c}{bandwidth}
&\multicolumn{1}{c}{bandwidth}
\\[1.0ex]
\hline
\noalign{\vspace{1.3ex}}
Pentium III (0.6 GHz)
&SDRAM PC133
& 32\hspace{0.3cm}
& 1.1
& 0.8
\\[1.3ex]
Athlon (1.2 GHz)
&DDRAM PC266
& 64\hspace{0.3cm}
& 2.1
& 1.3
\\[1.3ex]
Pentium 4 (1.4 GHz)
&RDRAM PC800
& 128\hspace{0.3cm}
& 3.2
& 2.1
\\[1.3ex]
\hline
\end{tabular}
\vspace{-0.16cm}
\end{table*}

Cache memories are expensive and usually too small to contain
a complete lattice Dirac field. The spinors thus tend to be 
overwritten before they are reused. 
It is possible, however, to increase the cache-hit probability
by dividing the lattice into small blocks that are visited one
after another. This technique (which is referred to as {\it strip-mining}\/)
is easy to implement and often
results in significant speed-up factors.

In many cases 
the cache-hit rate can also be enhanced by explicitly moving the data 
to the cache memory slightly before they are going to be used.
This can be achieved by inserting {\tt prefetch} instructions such as

\vspace{2ex}
{\tt
\noindent
\_\_asm\_\_ \_\_volatile\_\_ (\\[0.2ex]
\hbox{\hskip0.50cm}"prefetcht0 \%0"~::~"m" (*(address)));%
}

\vspace{2ex}
\noindent
at the appropriate places in the program.
The important point to note is that these
are executed ``out of order",
i.e.~while the arithmetic unit is busy on other data.
In ideal cases the memory latencies can thus be masked almost completely.

Data are moved from the main memory to the cache memory 
in blocks that are referred to as cache lines.
The cache line sizes and transfer rates for some of the recent
PC processors and memory types are listed in table~2.
When a prefetch instruction is issued, 
a cache line containing the byte at the specified address is moved to 
the cache memory.
Dirac spinors occupy one or more cache lines,
depending on the line size and the data alignment, 
and the programmer needs to take this into account
to make sure that all components are prefetched.

\subsection{Program portability}

Inline assembly and the {\tt SSE2} instructions
are system-specific and thus 
not as portable as an {\tt ISO} {\tt C} compliant 
code. It is conceivable that the {\tt C} language will be extended 
to include short vector data types once
the vector units become an industry standard,
and it would then be 
left to the compiler to produce assembler code that makes 
use of any advanced features of the processor.

For the time being, program portability can be 
preserved by first writing an {\tt ISO} {\tt C} program
for the specified task. Where this appears worth while,
inline assembly code may then be added that is conditionally compiled,
i.e.~only when the macro {\tt SSE2} for example
is defined. In this way it can be decided at 
compilation time whether some parts of the {\tt ISO} {\tt C} program
are to be replaced by system-specific code or not.

\subsection{FermiQCD}

Programs that make use of 
the {\tt SSE} unit along the lines described above
are now part of the 
FermiQCD package \cite{FermiQCDdoc}.
Many macros for SU(3) vector and matrix operations 
in the {\tt SSE} registers can be found there,
and the programs may also serve to illustrate
the use of prefetch instructions.
The package is freely accessible and can be downloaded
from a webpage at Fermilab \cite{FermiQCDweb}.

\section{NETWORK ISSUES}

\subsection{Commercial PC clusters}

In the course of the last two years or so,
fully configured rack-mounted clusters of PCs have become
a commodity. These machines usually have a switched network where 
any node can exchange data with any other node at full speed 
(see fig.~2). 

As will become clear below,
the bandwidth of the network is a critical parameter
for clusters that are intended for numerical simulations of lattice QCD.
The currently preferred network for such machines is
Myrinet$^{\rm TM}\!$ \cite{Myrinet}, which provides
a one-way bandwidth per channel of up to 250 Mbyte/s.
Switches for 8 to 128 nodes are available, and these can, in principle,
be connected to build clusters of any size.

\begin{figure}[tb]
\vspace{0.1cm}
\hspace{0.5cm}\includegraphics[scale=0.43]{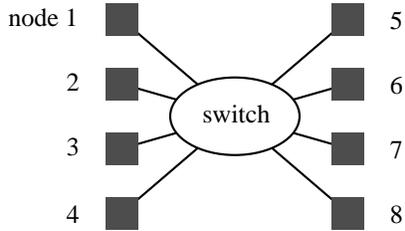}
\vspace{-0.4cm}
\caption{The nodes of a commercial PC cluster
are essentially complete PCs with one or two processors.
They are connected to
a switch that forwards data packages between them
as requested by the user programs.}
\label{fig:Cluster}
\vspace{-0.69cm}
\end{figure}

\subsection{How much bandwidth is needed?}

Instead of developing 
a general formula for the required bandwidth, 
it may be more instructive to consider a realistic study case.
So let us suppose that we have a machine with 32 nodes, each equipped
with 512~MB of memory and a processor that delivers a sustained
computational power of 1~Gflop/s in 32-bit arithmetic.
As for the network, we assume that it provides a simultaneous
bandwidth of 200 Mbyte/s per channel (the associated 
latencies are unimportant in the present context
since the data can be transferred in large packages).

Such a machine is sufficiently big for quenched QCD simulations 
with Wilson fermions on a
$96\times48^3$ lattice. We may, for example, distribute 
the lattice to the nodes in such a way that 
each node is associated with a sublattice of size 
$12\times12\times48^2$ (see fig.~3). 
The question is then how much time is required by
the communication overhead in the program that applies
the Wilson--Dirac operator to a given fermion field.

When the program runs through the lattice,
it reads the components of the input fermion field on the 
points of the local lattice and on all its boundary points.
The Dirac spinors residing on the latter
are stored in the memory of the logical neighbours of the current node
and need to be communicated through the network 
before the calculation starts.

If 32-bit arithmetic is used,
each Dirac spinor occupies 96 bytes of storage,
and the total amount 
of data that must be moved to every node is thus 
about 11~MB.
On our machine the time required for this is
\begin{eqnarray*}
  t_{\rm com}\kern-1.8em&&=2\times{96\;{\rm byte}\over 200\;{\rm Mbyte/s}}
              \times\frac{1}{3}V
  = 0.32\,\mu{\rm s}\times V,
\end{eqnarray*}
where $V$ denotes the volume of the local lattice.
Once all the data are available,
the application of the 
Wilson--Dirac operator takes about $1.4\,\mu$s per lattice point, and
the communication overhead is thus estimated to be 19\% of the total time.

\begin{figure}[t]
\vspace{0.0cm}
\hspace{1.9cm}\includegraphics[scale=0.24]{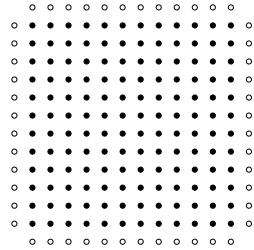}
\vspace{-0.5cm}
\caption{Cross-section of the local $12\times12\times48^2$ lattice
in the $(x_0,x_1)$ plane (full circles). The Di\-rac spinors on the 
$4\times12\times48^2$ boundary
points (open circles) need to be fetched from the neighbouring nodes.}
\label{fig:LocalLattice}
\vspace{-0.45cm}
\end{figure}

The result of this theoretical exercise shows
that current network bandwidths and processing speeds are not out
of balance. It should be noted, however, that the
ratio of the boundary to the bulk volume of the local lattice, and hence the 
communication time, 
tends to increase proportionally to the number of nodes
that are used for a given lattice.
In many cases it may, therefore, 
be more efficient to simulate several independent 
lattices in parallel.

\begin{table*}[htb]
\caption{PC clusters dedicated for lattice QCD simulations
(incomplete)}
\label{table:3}
\begin{tabular}{lcll}
\hline
\noalign{\vspace{1.0ex}}
\multicolumn{1}{c}{Institution}
&\multicolumn{1}{c}{\# nodes}
&\multicolumn{1}{c}{Processor(s)}
&\multicolumn{1}{c}{Status}
\\[1.0ex]
\hline
\noalign{\vspace{1.0ex}}
Fermilab
&$80$
&dual Pentium III (0.7~GHz)
&\hspace{1.0em}running
\\[1.0ex]
TC Dublin
&$32$
&dual Pentium III (1.0~GHz)
&\hspace{1.0em}running
\\[1.0ex]
MPI Munich
&$8$
&dual Pentium III (1.0~GHz)
&\hspace{1.0em}installation
\\[1.0ex]
DESY (Zeuthen)
&$16$
&dual Pentium 4 (1.7~GHz)
&\hspace{1.0em}ordered
\\[1.0ex]
DESY (Hamburg)
&$32$
&Pentium 4 (1.7~GHz)
&\hspace{1.0em}approved
\\[1.0ex]
Fermilab
&$\sim 150$
&dual Pentium 4
&\hspace{1.0em}approved
\\[1.0ex]
Jlab \& MIT
&$128$
&dual Pentium 4
&\hspace{1.0em}approved
\\[1.0ex]
\hline
\end{tabular}
\vspace{0.1cm}
\end{table*}

\subsection{Possible improvements}

PC processors are rapidly becoming faster, and it is clear that
the network performance will have to be improved
if the communication overhead is to remain small.
Increasing the one-way bandwidth or having 
simultaneous bi-directional data transfers is certainly an option.
With the current specification of the PCI bus (64 bit @ 66 MHz),
transfer rates of up to 500 Mbyte/s can perhaps be
reached.

An important reduction of the communication overhead 
might also be achieved 
if computation and communication can be made to overlap in time. 
The obvious difficulty here is  
that the PCI bus and the processor may be unable to access the memory 
concurrently with sufficient bandwidth, particularly so if 
the program is highly optimized.

\section{PC CLUSTERS FOR LATTICE QCD}

\subsection{Recent installations}

This year several dedicated machines
have been installed or will shortly be delivered
(see table~3). 
All these machines are commercial products that 
are equipped with Myrinet$^{\rm TM}\kern-2pt$
and come with the standard software environment. 

The clusters listed in the last two lines 
of table~3 are being funded through 
the {\it Scientific Discovery through Advanced Computing}\/
(SciDAC)
initiative of the US government \cite{SciDAC}. This programme 
is intended to help closing the gap
between nominal and actual performance of 
massively parallel computers and supports
``research and development on scientific modeling codes, as well
as on the mathematical and computing systems software that
underlie these codes".

\subsection{Do PC clusters scale to Tflop/s?}

Clusters with (say) 128 nodes currently 
reach sustained computational speeds of about 100--200 Gflop/s
in lattice QCD programs. 
A small farm of such machines thus delivers 
an integrated computational power in the Tflop/s range.
Having a farm of clusters is in fact not such an odd idea, since 
the total memory of each cluster will
in many cases be sufficiently big
to accommodate the lattice fields.
Statistics can then
be accumulated by running the same program 
(with different sequences of random numbers)
on several machines. 

\vfill\eject

Whether much larger numbers of processors can be integrated
in a single cluster in a useful way is not obvious, however,
and detailed studies will be needed
before a solid answer to this question can be given.
It seems likely, though, that 
the network switch will turn out to be a bottleneck
when the cluster exceeds a certain critical size.

A development project,
recently initiated at the University of Rome II and 
now supported by INFN,
tries to overcome this difficulty by arranging the 
nodes in a logical $n\times n$ matrix.
The nodes in each row and in 
each column are then connected to one another through 
$2n$ independent switches.
This network geometry naturally maps to the 
block divisions of the physical lattice that are usually adopted.
The routing of the data packages is thus simplified,
and switching latencies are expected to grow only relatively slowly
with the number of nodes.

\subsection{Uses of small machines}

When discussing computer performance and the need for
ever more powerful machines, 
an important fact is sometimes forgotten
that Giorgio Parisi pointedly described by \cite{Parisi}: 

\begin{center}
\vskip-0.6ex
\parbox[b]{7.2cm}{
\it ``In general a brute force approach to numerical simulations 
is not paying and a lot of ingenuity is needed in order to 
obtain the wanted results.''
}
\end{center}

\vskip-0.6ex
\noindent
Innovation in computational strategies and algorithms
is not only essential for progress in lattice QCD,
but it also represents an intellectual challenge and is 
certainly one of the driving forces in this field.
PC clusters are well suited to try out new physical
and algorithmic ideas,
because they are relatively easy to use and 
are in many cases sufficiently powerful for
significant tests to be performed in a reasonable time.
Even small machines can be very
useful for this kind of work
on which the large-scale projects eventually depend.

\begin{figure}[t]
\vspace{0.1cm}
\hspace{0.0cm}\includegraphics*[scale=0.54]{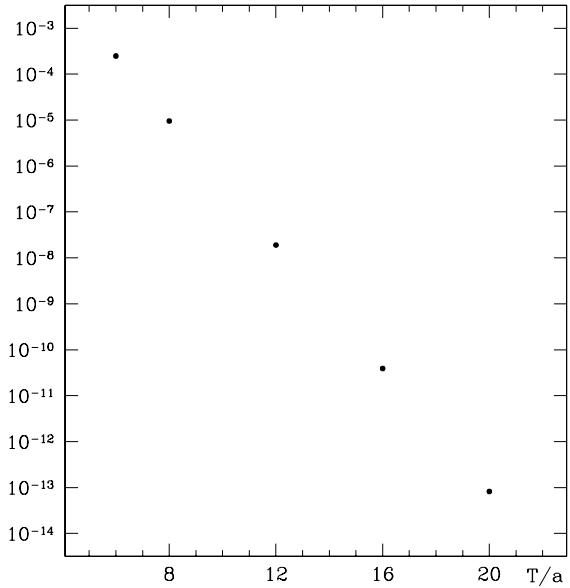}
\vspace{-1.1cm}
\caption{%
Polyakov loop correlation function at $\beta=5.7$ and 
distance $r=6a\simeq1.0\,\fm$, on a lattice of spatial size $L=12a$ 
and variable time extent $T$.
Statistical errors are not visible on this scale.}
\label{fig:CorrelationFunction}
\vspace{-0.5cm}
\end{figure}

For illustration, let us consider the
Polyakov loop two-point correlation function in the SU(3) gauge theory.
When the time extent $T$ of the lattice
is larger than (say) $1\,\fm$, the theory is in the confinement phase and
the correlation function at distance $r$ 
decays exponentially,
roughly like
\begin{eqnarray*}
  \langle P(r)^{\ast}P(0)\rangle\propto\rme^{-\sigma A},
  \qquad
  A=rT,
\end{eqnarray*}
where $\sigma\simeq1\,\GeV/\fm$ denotes the string tension.
Now if the established simulation algorithms are used,
the correlation function is obtained with 
statistical errors that tend to be essentially independent of $A$.
As a consequence the amount of computer time
needed for a specified relative precision is exponentially rising
and areas $A\geq 1\,\fm^2$ or so are practically unreachable.

It is clear that significant progress in this field 
can only be made if better simulation algorithms are found 
that lead to an exponential reduction of the statistical errors.
The recently proposed multi\-level algorithm \cite{Multilevel}
is of this type and works exceedingly well, at least
in the case of the Wilson plaquette action.

An impressive demonstration of this is given by 
the data plotted in fig.~4. At the chosen coupling, the 
lattice spacing $a$ is about $0.17\,\fm$, and the 
values of $A$ covered in the plot thus reach 
nearly $3.5\,\fm^2$. In spite of the fact that the signal decays
over many orders of magnitude in this range,
the equivalent of no more than about
$1000$ hours of processor time on a PC with 
$1.4$ GHz Pentium~4 processor 
was required 
to keep the statistical errors below 2\% at all values of $T$.

\vspace{2.0ex}

\section{CONCLUSIONS}

At this point it is quite clear that PC clusters are 
going to be widely used for lattice QCD simulations.
While smaller machines are ideal for development work and 
physics projects at an early stage,
large clusters or farms of clusters are certainly an option
for big installations, where the aim is to maximize the total
computational power for a given budget.

The cost per (32-bit sustained) Mflop/s
of a commercial PC cluster with Myrinet$^{\rm TM}\!$
network is currently \$\kern1pt3--4.
In terms of cost\--effectiveness, these machines are hence
very competitive.
How well PC clusters scale to large numbers of nodes, 
or whether it will be more useful to have 
a farm of smaller machines, remains to be seen.
Further experience with both the hardware
and the software will be needed, 
before it can be decided which is the best way to go.

Computers become more powerful at 
a breathtaking rate, but it seems unlikely that 
the hard problems in lattice QCD will gradually go away
as a result of this alone.
Improved numerical methods
(such as those reviewed by Mike Peardon \cite{Peardon})
are in fact often crucial for the success of a numerical 
simulation project. 

\vfill\eject

To explore a new approach
can be fairly pain\-ful, however, if there is no 
reasonably fast computer around that can be freely used.
PC~clus\-ters provide a good solution to this problem, and 
the fact that even small groups of lattice physicists may
be able to afford them opens the way for more people to 
contribute to the development of lattice QCD.

\vskip1ex
I am indebted to Roberto Petronzio and Jarda Pech for many
informative discussions on current network technologies.
Thanks also go to Filippo Palombi for 
communicating his benchmark
results for the Athlon processors and to
Robert Edwards, Karl Jansen, Paul Mackenzie, 
Mike Peardon, Massimo Di Pierro and Peter Wegner
for helpful correspondence.


\end{document}